\documentclass[showpacs,twocolumn,preprintnumbers,prl,amssymb,floatfix]{revtex4}
\usepackage{graphicx}
\usepackage{dcolumn}
\usepackage{bm}
\usepackage{amsmath}
\usepackage{amsfonts}
\usepackage{amssymb}
\usepackage{ulem}
\usepackage{color}

\begin{document}
\title{Two-dimensional Graphene with Structural Defects:\\
Elastic Mean Free Path, Minimum Conductivity and Anderson Transition}
\author{Aur\'{e}lien Lherbier$^{1,2}$}
\author{Simon M.-M. Dubois$^{1,3}$}
\author{Xavier Declerck$^{1,2}$}
\author{Stephan Roche$^{4,5}$}
\author{Yann-Michel Niquet$^{6}$}
\author{Jean-Christophe Charlier$^{1,2}$}
\affiliation{$^{1}$Universit\'e Catholique de Louvain (UCL), Institute of Condensed Matter and Nanoscience (IMCN), Place Croix du Sud 1 (NAPS-Boltzmann), 1348 Louvain-la-Neuve, Belgium}
\affiliation{$^{2}$European Theoretical Spectroscopy Facility (ETSF)}
\affiliation{$^{3}$University of Cambridge, Cavendish Laboratory, Theory of Condensed Matter group, JJ Thomson Avenue, Cambridge CB3 0HE, United-Kingdom}
\affiliation{$^{4}$Institut Catal\`{a} de Nanotecnologia (ICN) and CIN2, UAB Campus, E-08193 Barcelona, Spain}
\affiliation{$^{5}$Instituci\'o Catalana de Recerca i Estudis Avan{\c c}ats (ICREA), 08010 Barcelona, Spain}
\affiliation{$^{6}$CEA-UJF, INAC, SP2M/L\_Sim, 17 rue des Martyrs, 38054 Grenoble Cedex 9, France}

\date{\today}

\begin{abstract}
Quantum transport properties of disordered graphene with structural defects (Stone-Wales and divacancies) are investigated using a realistic $\pi-\pi^{*}$ tight-binding model elaborated from {\it ab initio} calculations. Mean free paths and semiclassical conductivities are then computed as a function of the nature and density of defects (using an order-N real-space Kubo-Greenwood method). By increasing of the defect density, the decay of the semiclassical conductivities is predicted to saturate to a minimum value of $4e^2/\pi h$ over a large range (\textit{plateau}) of carrier density ($>0.5\, 10^{14}\text{cm}^{-2}$). Additionally, strong contributions of quantum interferences suggest that the Anderson localization regime could be experimentally measurable for a defect density as low as $1\%$.
\end{abstract}


%
%
\maketitle

Clean graphene exhibits unique transport properties at the Dirac point. The density of charge carriers vanishes but the conductivity remains finite in the order of a few $e^{2}/h$~\cite{Geim,Castro}. This minimum of conductivity observed in the ballistic regime and related to contacts effects is also present in the diffusive regime where disorder plays an important role. In conventional two-dimensional disordered metals, it is well established that such a low conductivity leads to an Anderson-type insulator at low temperatures~\cite{Langendijk,Lee}. In the Anderson regime, the electronic states are spatially localized over a characteristic length scale, called localization length $\xi$~\cite{Lee}, which drives the exponential suppression of conductivity with system length. The theory of Anderson localization initially developed for electrons has been demonstrated to be ubiquitous in Physics, applicable to other types of particles, such as photons and atoms~\cite{Langendijk}. Surprisingly however, to date the observation of such metal-insulator transition remains elusive in graphene, even in low-mobility devices. One of the reasons stems from the peculiar nature of localization effects in graphene whose charge carriers behave as massless Dirac fermions with chirality degree of freedom at the origin of a sign reversal of the quantum correction to the semiclassical conductivity (weak antilocalization~\cite{McCann}). When intervalley scattering strongly predominates, ordinary weak localization is however expected to drive the system to an Anderson insulator~\cite{WAL,Lherbier}, but those theoretical results restrict to simplified disorder models, preserving the ${\rm sp}^{2}$ symmetry.

On the other hand, controlled defect engineering in ${\rm sp}^{2}$--carbon-based materials has become a topic of great excitment~\cite{Krasheninnikov}. Indeed, the electronic (and transport) properties of carbon nanotubes~\cite{Charlier} and graphene-based materials~\cite{Suenaga,Cresti} can be considerably enriched by chemical modifications, including molecular doping and functionalization. Strong modifications of ${\rm sp}^{2}$-bonded carbon materials by incorporation of ${\rm sp}^{3}$ defects is also a suitable route for making graphene more sensitive to localization phenomena. Recently an attempt to turn graphene into a true band insulator was endeavored by transforming ${\rm sp}^{2}$ bonds into ${\rm sp}^{3}$ by hydrogenation~\cite{Elias}. Similar results have been experimentally debated in fluorinated~\cite{Withers} or ozone treated graphene flakes~\cite{Moser}. Using ion irradiation, specific types of structural defects (vacancies) can be introduced in ${\rm sp}^{2}$-based carbon nanostructures. Convincing room-temperature signatures of an Anderson regime in ${\rm Ar}^{+}$ irradiated carbon nanotubes have been reported~\cite{Biel1}. In contrast, the conductivity of irradiated two-dimensional graphene saturates at the Dirac point above $e^{2}/h$ even down to cryogenic temperatures~\cite{Chen}, suggesting a strong robustness of defective graphene.

In this Letter, the electronic and transport properties of disordered graphene are theoretically explored by introducing structural defects (Stone-Wales and divacancies) randomly distributed in the honeycomb lattice. The presence of these structural defects triggers resonant impurity levels, which broaden and generate impurity bands as their density is increased. {\it Ab initio} calculations are performed to accurately describe the local energetics around the defects, and a tight-binding (TB) model ($\pi$-$\pi^{*}$) is elaborated from the corresponding band structures. Additionally, an efficient order-N real-space Kubo-Greenwood transport method is used to follow the transition from the diffusive to the insulating regime.

\begin{figure*}[ht!]
\begin{center}
\leavevmode
\includegraphics[width=1.0\textwidth]{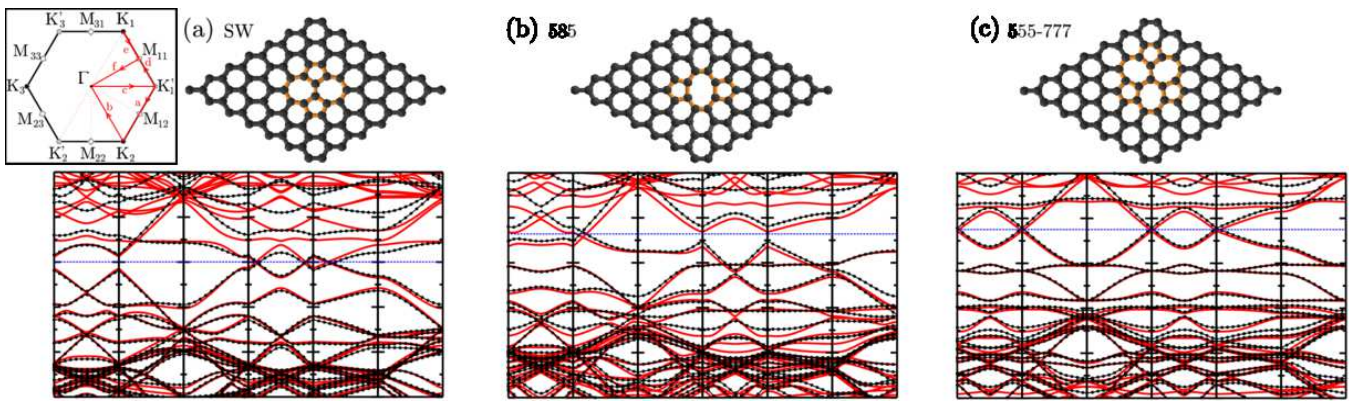}
\caption{(color online). Top panel : Schematic of various defects: Stone-Wales (a), 585 (b) and 555-777 (c) divacancies. Bottom panel: Corresponding electronic band structures computed using the SIESTA package (red lines) and parametrized using a $3^{\rm{rd}}$ nearest-neighbors TB $\pi$-$\pi^{*}$ model (black dotted lines) along high-symmetry lines in the Brillouin zone (inset). Fermi energy is set at zero and the Dirac energy is indicated with a horizontal blue dashed line.}
\label{fig1}
\end{center}
\end{figure*}

In the present study, three types of non-magnetic structural defects are considered. The first one is the so-called Stone-Wales (SW) defect which consists of two heptagons connected with two pentagons (Fig.\ref{fig1}.a). Although its presence has already been experimentally reported in graphene~\cite{Suenaga}, its influence on the transport properties remains unknown. Ma \textit{et al.} have reported in a theoretical study\cite{Krasheninnikov} that a SW defect could produce a slight out-of-plane deformation. We consider here an in-plane geometry as a first approach. The two other defects consist in various reconstructions of the divacancy : one being associated to the formation of 2 pentagons and 1 octagon (585 - Fig.\ref{fig1}.b), while the other is composed of 3 pentagons and 3 heptagons (555-777 - Fig.\ref{fig1}.c). Our {\it ab initio} calculations predict the 555-777 reconstruction to be more stable than the 585 one by $\sim0.9$eV, in agreement with previous theoretical predictions~\cite{GundoLee}. Finally, divacancies are known to be more stable than two isolated monovacancies whose migration energy barrier is rather low~\cite{Krasheninnikov,GundoLee}. In contrast to the monovacancies, the divacancies do not require a spin-dependent electronic structure treatment~\cite{Palacios}.

To reduce the computational cost of the transport calculations, the electronic band structures (Fig.\ref{fig1}) first computed with the SIESTA code~\cite{SIESTA,Siesta_comput} are then reproduced using a parametrized $3^{\rm{rd}}$ nearest-neighbors $\pi$-$\pi^{*}$ TB Hamiltonian model~\cite{3nnTBmodel,Lenosky}. The agreement between this TB model and the {\it ab initio} calculation is quite satisfactory, especially in the transport energy region $[-1,1]$eV around the Dirac point. The observed differences at higher energies are basically related to the inability of the $\pi$-$\pi^{*}$ TB model to reproduce the conduction bands of graphene along K-M branch. The SW defect does not display any doping character since both Fermi and Dirac energies are aligned (Fig.\ref{fig1}.a). On the contrary, both divacancies are found to act as acceptors since Fermi energy is below the Dirac point (Fig.\ref{fig1}.b-c).

\begin{figure}[t]
\begin{center}
\leavevmode
\includegraphics[width=0.83\columnwidth]{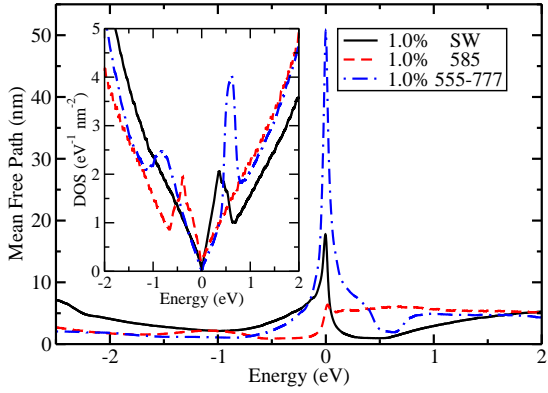}
\caption{(color online). Elastic mean free paths (main frame) and total DOS (inset) for disordered graphene with a defect density of $1\%$ of SW (solid line), 585 (dashed line) or 555-777 (dotted dashed line) divacancies. All curves have been aligned to make coincide the minimum of DOS with the zero energy.}
\label{fig2}
\end{center}
\end{figure}

Based on the present TB model, the density of states (DOS) are calculated using the recursion method, which is very efficient for large disordered systems~\cite{Kubo1}. In Fig.\ref{fig2} (inset), the total DOS corresponding to a defect density ($n_{d}$) of 1$\%$ are displayed for the three different defects. A large difference in the position of the defect-induced resonant impurity levels can be observed. For instance, the quasi-bound states localized around the SW defect yields a bump in the DOS around $0.35$eV above the Dirac energy (Fig.\ref{fig2} -- solid line). The 585 defect exhibits a similar DOS fingerprint but with an opposite behaviour to the case of SW defect. Indeed, for this defect the corresponding peak appears around $-0.35$eV (Fig.\ref{fig2} -- dashed line). Finally, in presence of 555-777 defects, two different DOS impurity-peaks are reported : a wider impurity band in the hole region ($-0.8$eV) and a sharp peak on the electron side ($0.6$eV) (Fig.\ref{fig2} -- dashed-dotted line). In all cases, the intensity and the width of the defect-induced peaks are found to increase with $n_{d}$.

Transport properties of large and disordered graphene systems are then calculated using an efficient order-N Kubo-Greenwood method~\cite{Kubo1}. This powerful technique gives a direct access to the elastic mean free paths at energy $E$, extracted from the wavepacket dynamics. The latter is characterized by the time-dependent diffusivity coefficient $D(E,t)=\Delta R^2(E,t) / t $ with $\Delta R^2 = \Delta X^2 + \Delta Y^2 $ and $\Delta X^2 (E,t) = \rm{Tr}[\delta (E-\hat{H}) \vert \hat{X}(t) - \hat{X}(0)\vert^2] / \rm{Tr}[\delta (E-\hat{H})]$. $\rm{Tr}$ is the trace over $p_z$ orbitals and $\rm{Tr}[\delta (E-\hat{H})] / S = \rho(E)$ is the total DOS (per unit of surface). The two position operators $\hat{X}(t)$ and $\hat{Y}(t)$ are expressed in the Heisenberg representation ($\hat{X}(t) = U^{\dagger}(t)X(0)U(t)$) and the time evolution operator $\hat{U}(t)=\Pi_{n=0}^{N-1}\exp(i\hat{H}\Delta t/\hbar)$ with $\Delta t$ the chosen time step, is computed with a Chebyshev polynomial expansion method~\cite{Kubo1}. Calculations are performed for several initial random phase wavepackets, and for total elapsed time $t \approx 1.5$ps. The typical size of the simulated system is $\sim0.074 \mu \text{m}^{2}$ (containing $2.8\times10^6$ carbon atoms), large enough to avoid finite size effects. In the Kubo formalism, the different transport regimes can be inferred from the behaviour of $D(E,t)$. The wavepacket velocity $v(E)$ can be extracted from the short time behavior of the diffusivity, $D(E,t) \sim v^2(E)t$, while the elastic mean free path $\ell_{e}(E)$ is estimated from the maximum of the diffusivity, $D_{\text{max}}(E)=2v(E)\ell_{e}(E)$. Finally the Kubo semiclassical conductivity reads $\sigma_{sc}(E)=\frac{1}{4}e^{2}\rho(E)D_{\text{max}}(E)$.

The mean free paths are calculated for defect densities varying from $n_{d}=0.1\%$ to $1\%$. At a given energy, $\ell_{e}$ is predicted to be inversely proportional to $n_{d}$, as expected from the Fermi golden rule. $\ell_{e}$ also displays a strong energy dependence, with dips around the Dirac point where it reaches values as low as few nanometers. These dips are associated with the resonant impurity states, which are specific to each defect. $\ell_{e}(E)$ can change by one order of magnitude around the Dirac point depending on the nature of the defect and the energy of charge carriers, as shown on Fig.\ref{fig2} (main frame) for a defect density $n_{d}=1\%$. For instance, $\ell_{e}$ is estimated to be $\sim5$nm at the Dirac point in presence of 585 defects, whereas $\ell_{e}\simeq 50$nm for the 555-777 defects because the two associated energy resonances ($-0.8$ and $0.6$eV) are farther from the Dirac point. Consequently, this defect nature dependence will strongly impact both semiclassical and quantum transport regimes, as discussed below.

\begin{figure}[t]
\begin{center}
\leavevmode
\includegraphics[width=\columnwidth]{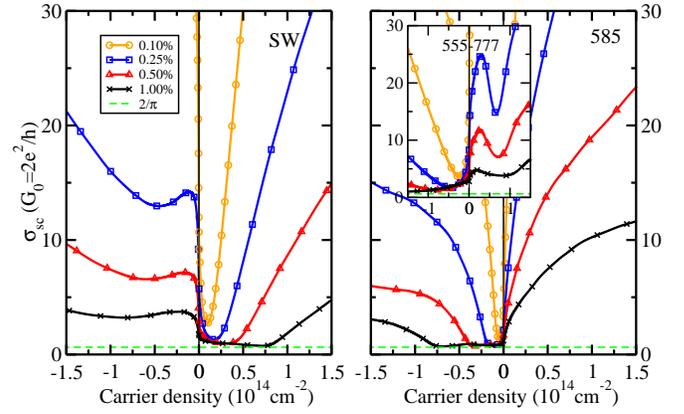}
\caption{(color online). Semiclassical conductivities in disordered graphene versus carrier densities and various defect densities (from 0.1$\%$ to $1\%$) of Stone-Wales (left panel), 585 (right panel main frame) and 555-777 (right panel inset).}
\label{fig3}
\end{center}
\end{figure}

Figure~\ref{fig3} illustrates the semiclassical conductivities for disordered graphene with the three possible defects evaluated as $\sigma_{sc}(n)=\frac{1}{2}e^{2}\rho(n)v(n)\ell_{e}(n)$, where $n(E)$ is the carrier density~\cite{precision_Fig3_n(E)}. In addition to the strong dependence of $\sigma_{sc}$ on the carrier energy, a saturation of the conductivity decay with $n_{d}$ is observed, reaching the minimal value of $\sigma_{sc}^{\rm min}= 4e^{2}/\pi h$. As $n_{d}$ increases, a large plateau of minimum conductivity develops close to the Dirac point, similarly to the case of simpler point defects~\cite{Wehling}. The origin of such a minimum of conductivity has recently been strongly debated in the literature~\cite{Minimcond}. Our simulations suggest that such a value could be associated to a graphene system containing structural defects, in absence of quantum interferences (QI). One further observes that for each type of defects, the plateaus of minimal conductivity extend over the energy window encompassing the corresponding resonant impurity bands. Indeed, for the SW case, this plateau is primarily located at energies above the Dirac point, in contrast with the 585 case, where the minimum conductivity is observed below the Dirac point. The case of 555-777 is more complicated due to the presence of two impurity resonant energy windows. 
In the present study, a \textit{rigid defect model} has been considered, ignoring the screening effects which could occur at high carrier densities. However, we have carefully checked that increasing the carrier density in the {\it ab initio} calculations does not significantly modify the bandstructure, supporting this \textit{rigid defect model} approximation.

\begin{figure}[t]
\begin{center}
\leavevmode
\includegraphics[width=0.85\columnwidth]{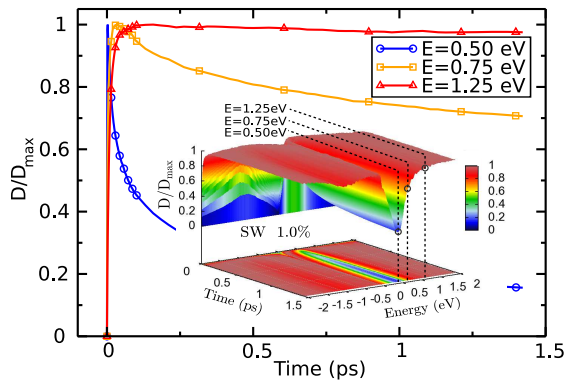}
\caption{(color online). Time dependent diffusion coefficient $D(E,t)$ normalized to $D_{\max}=2v\ell_{e}$ at three selected energies and for a $1\%$ density of SW defects. Inset: Enlarged view of $D(t)/D_{\rm max}$ over the energy spectrum.}
\label{fig4}
\end{center}
\end{figure}

One noteworthy observation is that the predicted short mean free paths close to the Dirac point favors strong contributions of QI, as long as the transport regime remains quantum coherent. In the present Kubo formalism, these QI contributions can be evidenced in the ratio $D(t)/D_{\rm max}$ which departs from unity at long times in the presence of QI. Fig.\ref{fig4} (main panel) shows several typical behaviors of $D(t)/D_{\rm max}$ at selected energies and for the case of $n_{d}=1\%$ of SW defects. A more global picture (over a larger part of the energy spectrum) is also given in the inset. At $E=0.5$eV (close enough to the resonance energy associated to the SW, see Fig.2 (inset)), $D(t)/D_{\rm max}$ exhibits a fast decay consistent with the estimated short localization length ($\xi\sim10$nm). Indeed, following the scaling theory of localization, $\xi$ can be determined once the semiclassical transport length scales are known, as $\xi(E)=\ell_{e}(E)\exp(\pi h\sigma_{sc}(E)/2e^{2})$~\cite{Lee}. Similar values are found at the resonant energies for both $1\%$ of 585 defects ($\xi\sim10$nm) and  for $1\%$ of 555-777 defects ($\xi\sim25$nm). Actually, in the energy windows where $\sigma_{sc}$ saturates to $\sigma_{sc}^{\rm min}$, short localization lengths ($\xi<100$nm) are obtained for all defect cases. Much longer $\xi$ are predicted for energies outside the minimum conductivity plateaus, as illustrated in Fig.4 for $E=1.25$eV where the contribution of QI can even become vanishingly small.

Specific conductance fingerprints will be thus obtained depending on the nature and density of defects. Although real samples of defective (irradiated) graphene are likely to encompass a mixture of those different defects, the saturation of the semiclassical conductivities and the typical localization lengths (in the range of $30$nm for $1\%$ defects) close to the Dirac point will be a robust common feature to all possible cases. Considering the reported weak electron-phonon coupling\cite{Morozov}, such insulating state should be observable even at room temperatures (using suspended graphene preferentially).

J.-C.C. and A.L. acknowledge financial support from the FNRS of Belgium. Parts of this work are connected to the Belgian Program on Interuniversity Attraction Poles (PAI6), to the NanoHymo ARC, to the ETSF e-I3 project (grant n.$^{\circ}$ 211956), and to the NANOSIM-GRAPHENE Project No. ANR-09-NANO-016-01. Computational resources are provided by the UCL-CISM.


\begin{thebibliography}{50}
\bibitem{Siesta_comput}
The DFT calculations are performed using SIESTA~\cite{SIESTA} within the GGA for the exchange-correlation functional. Troullier-Martins pseudopotentials are used to account for the core electrons~\cite{TM}. The valence electron wave functions are expanded in a double-$\zeta$ polarized basis set of finite-range pseudoatomic orbitals.
\bibitem{precision_Fig3_n(E)}
In the text, carrier density $n(E)$ is evaluated by integrating the DOS from the Dirac point to the energy $E$.
\end{thebibliography}
\end{document}